\documentclass[times,twocolumn,final,authoryear]{elsarticle}

\usepackage{ycviu}
\usepackage{framed,multirow}

\usepackage{amssymb}
\usepackage{latexsym}

\definecolor{newcolor}{rgb}{.8,.349,.1}
\usepackage{hyperref}       
\usepackage{url}            
\usepackage{booktabs}       
\usepackage{amsfonts}       
\usepackage{nicefrac}       
\usepackage{microtype}      
\usepackage{lipsum}
\usepackage{pifont}
\usepackage{amsmath}
\usepackage{graphicx}
\graphicspath{ {./images/} }
\usepackage{amsfonts}       
\usepackage{circledsteps}
\usepackage{subcaption}
\usepackage{color, colortbl}
\usepackage[normalem]{ulem}
\usepackage{multirow}
\usepackage{bbm}

\useunder{\uline}{\ul}{}
\definecolor{Gray}{gray}{0.9}

\newcommand{\xa}{\textcolor{red}{\mathbf{x}^{(a)}}}

\newcommand{\xaq}{\textcolor{red}{\mathbf{x}_q^{(a)}}}
\newcommand{\xac}{\textcolor{red}{\mathbf{x}_c^{(a)}}}

\newcommand{\xaqrec}{\textcolor{red}{\mathbf{\hat{x}}^{(a)}_{q}}}

\newcommand{\za}{\textcolor{red}{\mathbf{z}^{(a)}}}
\newcommand{\tildza}{\textcolor{red}{\tilde{\mathbf{z}}^{(a)}}}
\newcommand{\zaG}{\textcolor{red}{\mathbf{w}^{(a)}}}
\newcommand{\zaGN}{\textcolor{red}{\mathbf{w}^{'(a)}}}

\newcommand{\xv}{\textcolor{blue}{\mathbf{x}^{(v)}}}

\newcommand{\xvq}{\textcolor{blue}{\mathbf{x}_q^{(v)}}}
\newcommand{\xvc}{\textcolor{blue}{\mathbf{x}_c^{(v)}}}

\newcommand{\xvqrec}{\textcolor{blue}{\mathbf{\hat{x}}^{(v)}_{q}}}

\newcommand{\zv}{\textcolor{blue}{\mathbf{z}^{(v)}}}
\newcommand{\tildzv}{\textcolor{blue}{\tilde{\mathbf{z}}^{(v)}}}
\newcommand{\zvG}{\textcolor{blue}{\mathbf{w}^{(v)}}}
\newcommand{\zvGN}{\textcolor{blue}{\mathbf{w}^{'(v)}}}

\newcommand{\x}{\mathbf{x}}

\journal{Computer Vision and Image Understanding}

\begin{document}

\thispagestyle{empty}

\clearpage
\thispagestyle{empty}

\ifpreprint
  \vspace*{-1pc}
\else
\fi

\ifpreprint
  \setcounter{page}{1}
\else
  \setcounter{page}{1}
\fi
\clearpage

\clearpage

\begin{frontmatter}

\title{A vector quantized masked autoencoder for audiovisual speech emotion recognition}

\author[]{Samir \snm{Sadok}\corref{cor1}} 
\cortext[cor1]{Corresponding author. Now at INRIA, Univ. Grenoble Alpes, CNRS, LJK }
\ead{samir.sadok@centralesupelec.fr}
\author[]{Simon \snm{Leglaive}}
\author[]{Renaud \snm{S\'eguier}}

\address{CentraleSupélec, IETR UMR CNRS 6164, France}


\begin{abstract}
An important challenge in emotion recognition is to develop methods that can leverage unlabeled training data. In this paper, we propose the VQ-MAE-AV model, a self-supervised multimodal model that leverages masked autoencoders to learn representations of audiovisual speech without labels. The model includes vector quantized variational autoencoders that compress raw audio and visual speech data into discrete tokens. The audiovisual speech tokens are used to train a multimodal masked autoencoder that consists of an encoder-decoder architecture with attention mechanisms. The model is designed to extract both local (i.e., at the frame level) and global (i.e., at the sequence level) representations of audiovisual speech. During self-supervised pre-training, the VQ-MAE-AV model is trained on a large-scale unlabeled dataset of audiovisual speech, for the task of reconstructing randomly masked audiovisual speech tokens and with a contrastive learning strategy. During this pre-training, the encoder learns to extract a representation of audiovisual speech that can be subsequently leveraged for emotion recognition. During the supervised fine-tuning stage, a small classification model is trained on top of the VQ-MAE-AV encoder for an emotion recognition task. The proposed approach achieves state-of-the-art emotion recognition results across several datasets in both controlled and in-the-wild conditions.
\\\\

\end{abstract}



\end{frontmatter}


\section{Introduction}
\label{sec:intro}

Emotions are primarily communicated through nonverbal cues, such as tone, voice, facial expressions, and body language, rather than the words we use in oral communication \citep{mehrabian2017nonverbal}. With the increasing prevalence of audiovisual interfaces, there is a growing demand for systems capable of accurately recognizing emotions from audiovisual speech signals. Indeed, combining audio and visual modalities has proven to be highly effective in various tasks, in particular speech emotion recognition (SER) \citep{el2011survey, gao2019co, zhao2019sound, sadok2024multimodal, ramachandram2017deep, tsai2019multimodal, schoneveld2021leveraging}.

Supervised learning using labeled datasets is the dominant machine learning paradigm in emotion recognition. However, obtaining datasets with emotion labels is often resource-intensive and impractical to scale. In fact, many emotion recognition datasets rely on actors simulating emotions, which demands significant time and effort during data collection. Another strategy consists of manually collecting and annotating in-the-wild data. However, emotion labeling is an ambiguous task, and annotators may not reach a consensus \citep{busso2008iemocap}. 

Therefore, one important challenge is to develop methods that can leverage unlabeled training data. Self-supervised learning (SSL) has recently emerged as a promising technique to address this challenge \citep{wang2021fine, dib2023s2f2, chen2023exploring, liu2022audio, zhang2022survey}. SSL models are pre-trained on a large-scale unlabeled dataset to solve a pretext task, and they are subsequently fine-tuned on a small amount of labeled data to solve a given task~\citep{gong2022ssast, pepino2021emotion, jegorova2023ss}.

In this paper, we present the first multimodal SSL approach based on masked autoencoders \citep{he2022masked} for emotion recognition. We propose the VQ-MAE-AV model, a vector quantized (VQ) masked autoencoder (MAE) designed for audiovisual (AV) speech representation learning and applied to emotion recognition. 
The proposed approach first involves quantizing the audio and visual speech modalities to create discrete tokens, using vector quantized variational autoencoders (VQ-VAEs). Then, we pre-train the VQ-MAE-AV model in a self-supervised manner to reconstruct audiovisual speech tokens from partially visible inputs, and with an additional contrastive learning strategy. The VQ-MAE-AV model is based on an encoder-decoder architecture with attention mechanisms to fuse the modalities. Its self-supervised pre-training leverages 1~000 hours of audiovisual speech without needing emotion labels, allowing the VQ-MAE-AV model to learn an internal representation of audiovisual speech. Finally, this learned representation is used as input for an emotion recognition model trained on small-scale audiovisual speech datasets labeled with emotion categories.
The experimental results demonstrate that the proposed VQ-MAE-AV model achieves superior performance compared to state-of-the-art methods across several datasets in both controlled and in-the-wild conditions. Additionally, extensive ablation experiments are presented to investigate the impact of different model designs. 

The paper is organized as follows. In Section~\ref{sec:related_work} we present the related work. The proposed VQ-MAE-AV model is introduced in Section~\ref{sec:VQ-MAE-AV}. Experiments are presented in Section~\ref{sec:Experiments} and we conclude in Section~\ref{sec:conclusion}. The code and qualitative results are available at
\href{https://samsad35.github.io/VQ-MAE-AudioVisual}{https://samsad35.github.io/VQ-MAE-AV}.

\begin{figure*}[ht]
    \centering
     \begin{subfigure}[b]{0.99\textwidth}
         \centering
         \includegraphics[width=\textwidth]{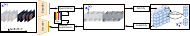}
        \caption{Discrete visual tokens.}
        \label{fig:visual-tokens}
     \end{subfigure}
     \hfill
     \begin{subfigure}[b]{0.99\textwidth}
         \centering
        \includegraphics[width=\textwidth]{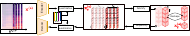}
        \caption{Discrete audio tokens.}
        \label{fig:audio-tokens}
     \end{subfigure}
        \caption{Discrete audio and visual tokens creation: (i) fully-convolutional VQ-VAEs are trained independently on the audio and visual modalities (see Section~\ref{subsec:VQVAE}); (ii) discrete audio and visual tokens are built from the quantized representations provided by the frozen VQ-VAE encoders (see Section~\ref{subsec:discrete_tokens}).}
        \label{fig:vqvae}
\end{figure*}

\section{Related work}
\label{sec:related_work}

\paragraph{Self-supervised learning approaches} 

SSL models can be broadly classified into two main categories: discriminative and generative approaches \citep{zhang2022survey}. Discriminative SSL focuses on creating pairs or groups of data samples and formulating loss functions that allow the model to differentiate or group these samples, which can later benefit downstream tasks~\citep{chen2020improved, chen2020simple}. For instance, pretext tasks can consist of solving jigsaw puzzles~\citep{noroozi2016unsupervised} or predicting image rotations~\citep{gidaris2018unsupervised}. Contrastive learning has emerged as the predominant paradigm in discriminative SSL~\citep{alayrac2020self, akbari2021vatt}.
In contrast, generative SSL involves generating or reconstructing segments of unlabeled and potentially corrupted data using an encoder-decoder model~\citep{he2022masked, bao2021beit, xie2022simmim}. The latent representation produced by the encoder can then be leveraged for downstream tasks.

\paragraph{Masked autoencoder} 

The present paper focuses on the MAE, an SSL generative model that employs an asymmetric encoder-decoder architecture with input masking~\citep{he2022masked}. The MAE approach is inspired by the concept of masked language modeling~\citep{devlin2018bert} and has been successfully applied to image modeling thanks to the development of the Vision Transformer (ViT)~\citep{dosovitskiy2020image}. The MAE has recently been extended to audio using a 2D time-frequency representation~\citep{gong2022ssast, baade2022mae, xu2022masked}. In the MAE paradigm, the input is divided into non-overlapping patches, each represented by a token embedding. Some tokens are masked, typically 75\% for image/audio modeling and 15\% for text modeling, and only the visible tokens are fed to the encoder. The encoder outputs are used to reconstruct the masked tokens through a lightweight decoder. To successfully reconstruct the masked tokens, the encoder must capture a semantic representation, which can then be effectively transferred to downstream tasks~\citep{he2022masked}. It was recently shown that combining the task of reconstructing masked tokens with contrastive learning can improve the representation learned by an MAE~\citep{huang2022contrastive, gong2022ssast}. 

\paragraph{Extensions of the MAE for sequential and multimodal representation learning} 

A recent extension of the MAE was presented in~\citep{feichtenhofer2022masked, tong2022videomae} for modeling image sequences, called Video-MAE. This method employs the same architecture as the vanilla MAE \citep{he2022masked} but incorporates a masking process from video ViT (ViViT)~\citep{arnab2021vivit}. Since videos often contain redundant information, particularly in scenes with no motion, the authors proposed a cubic masking approach along the temporal dimension, combined with a high masking ratio of 90\%.
Other works have extended the MAE to handle multimodal data~\citep{bachmann2022multimae, geng2022multimodal, gong2022contrastive}. MultiMAE~\citep{bachmann2022multimae} encodes a small random subset of visible tokens from multiple modalities (RGB, depth, and semantic images) and trains the model to reconstruct the missing tokens. M3AE~\citep{geng2022multimodal} is a unified MAE architecture for two input modalities (image and text). The main difference between M3AE and MultiMAE lies in the architecture of the decoder. The MulitMAE approach introduces individual decoders for each modality, enhancing their fusion by incorporating a cross-attention layer at the beginning of each decoder. On the other hand, the M3AE approach uses a single decoder that takes the concatenated tokens from all modalities as input.

\paragraph{Adaptation and improvement of the MAE} 

In the literature, MAEs are typically trained using the \emph{L1} or \emph{L2} losses, which can negatively affect the reconstruction quality of the masked tokens, resulting, for instance, in blurred or noisy images or sounds. Studies have demonstrated that improving the quality of MAE reconstructions can be beneficial in terms of downstream task performance~\citep{he2022masked}. Several approaches have been proposed for that purpose, for instance adding a perceptual loss~\citep{dong2021peco} or using discrete representations obtained from VQ generative adversarial networks (VQ-GANs)~\citep{esser2021taming} or variational autoencoders (VQ-VAEs)~\citep{van2017neural} to train the MAE~\citep{li2022mage, sadok2023vector}. While these works only considered a unimodal setting, the present paper proposes a multimodal MAE for audiovisual speech representation learning. 
Recently, a model called CAV-MAE~\citep{gong2022contrastive} was proposed combining MAEs and contrastive learning to learn a representation from audiovisual data. However, unlike the proposed VQ-MAE-AV model, CAV-MAE processes raw audiovisual data instead of vector-quantized latent representations, employs simple mean pooling strategies rather than attention pooling, and has not been applied to SER.

\begin{figure*}[t]
    \centering
    \includegraphics[width=\textwidth]{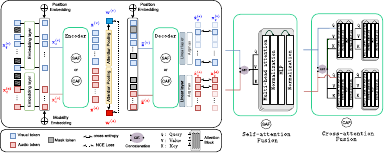}
    \caption{VQ-MAE-AV model structure. See the first paragraph of Section~\ref{sec:VQ-MAE-AV} for a complete description of the pipeline.}
    \label{fig:Overview}
\end{figure*}

\section{The VQ-MAE-AV model}
\label{sec:VQ-MAE-AV}

This section presents the VQ-MAE-AV model, which is illustrated in Fig.~\ref{fig:vqvae} and \ref{fig:Overview} and can be summarized as follows:
\begin{itemize}
\setlength{\itemindent}{0in}
    \item Fully-convolutional VQ-VAEs are trained independently on the audio and visual modalities (see Section~\ref{subsec:VQVAE});
    \item Discrete audio and visual tokens are built from the quantized representations provided by the frozen VQ-VAE encoders (see Section~\ref{subsec:discrete_tokens});
    \item A proportion of the discrete audio and visual tokens is masked out, using a coupled masking strategy between the two modalities (see Section~\ref{subsec:masking});
    \item The visible audio and visual tokens are replaced with trainable continuous embedding vectors (see Section~\ref{subsec:embedding}), which are fed to the VQ-MAE-AV encoder (see Sections~\ref{subsubsec:attention_block} and \ref{subsubsec:encoder}, where we present two strategies based on attention mechanisms to fuse the modalities);
    \item Attention pooling is used to compute global sequence-wise tokens that are specific to each modality (see Section~\ref{subsubsec:global_token});
    \item The token-wise representation obtained from the encoder is combined with mask tokens and fed to the VQ-MAE-AV decoder, which tries to reconstruct the original discrete audio and visual tokens (see Section~\ref{subsubsec:decoder});
    \item The VQ-MAE-AV model is trained in a self-supervised manner to minimize (i) the cross-entropy loss between the reconstructed and original tokens and (ii) a contrastive loss between the audio and visual global tokens (see Section~\ref{subsec:loss_function});
    \item After self-supervised learning, a small classification model is trained and the VQ-MAE-AV encoder is fine-tuned for supervised audiovisual SER (see Section~\ref{subsec:finetuning}).
\end{itemize}
This section will present each above-listed aspect of the model in more detail.

\subsection{Vector quantized variational autoencoder}
\label{subsec:VQVAE}

The proposed multimodal self-supervised approach uses the discrete latent representation of two pre-trained and frozen VQ-VAEs \citep{van2017neural}. Specifically, as illustrated in Fig.~\ref{fig:vqvae}, we use the VQ-VAE-A (A for audio) and VQ-VAE-V (V for visual) encoders to obtain compressed and quantized representations of the input speech power spectrogram $\xa \in \mathbb{R}^{T_a \times D}$ and of the input image sequence $\xv \in \mathbb{R}^{T_v \times H \times W \times C}$, where $T_a$ and $D$ correspond to the time and frequency dimensions of the audio modality, and $T_v$, $H$, $W$ and $C$ correspond to the time, height, width, and channel dimensions of the audio modality. The audio and visual quantized representations are denoted by $\xaq \in \mathbb{N}^{T_a \times D'}$ and $\xvq \in \mathbb{N}^{T_v \times H' \times W'}$, respectively. Each entry of $\xaq$ and $\xvq$ corresponds to the index of a vector in the VQ-VAE codebooks.
Notably, $\xaq$ retains the time-frequency structure of the original spectrogram, while $\xvq$ retains the spatio-temporal structure of the original sequence images. This is because the VQ-VAE-A and VQ-VAE-V models are designed to be fully convolutional on the frequency and spatial axes, respectively, and they process the frames within a sequence independently. Therefore, compression occurs along the frequency axis ($D' \ll D$) for $\xaq$ and along the x and y-axes of the image ($H' \ll H$, $W' \ll W$) for $\xvq$.
As shown in Fig.~\ref{fig:Overview} and discussed in the following subsections, the proposed MAE-based self-supervised learning approach operates on these discrete and compressed representations before audiovisual speech reconstruction using the VQ-VAE decoders. 

The training procedure of the VQ-VAEs follows the original approach presented in \cite{van2017neural}. In particular, the VQ-VAE loss functions involve a reconstruction term between the original and reconstructed data, which corresponds to the mean squared error for the visual modality and to the Itakura-Saito divergence for the audio modality \citep{fevotte2009nonnegative}. More details are provided in Section~\ref{sec:architecture}. 

\subsection{Discrete audio and visual tokens}
\label{subsec:discrete_tokens}

As shown in Fig.~\ref{fig:vqvae}, the audio and visual quantized representations $\xaq \in \mathbb{N}^{T_a \times D'}$ and $\xvq \in \mathbb{N}^{T_v \times H' \times W'}$ from the output of the VQ-VAE encoders are divided into non-overlapping patches to build discrete tokens  $\xaq \in \mathbb{N}^{(n_{t_a} \cdot t_a) \times (n_d \cdot d)}$ and  $\xvq \in \mathbb{N}^{(n_{t_v} \cdot t_v) \times (n_h \cdot h) \times (n_w \cdot w)}$, where $T_a = n_{t_a} \cdot t_a$, $D' = n_d \cdot d$, $T_v = n_{t_v} \cdot t_v$, $H' = n_h \cdot h$, and $W' = n_w \cdot w$.
These representations are reshaped to $\xaq \in \mathbb{N}^{(n_{t_a} \cdot n_d) \times (t_a \cdot d)}$ and $\xvq \in \mathbb{N}^{(n_{t_v} \cdot n_h \cdot n_w) \times (t_v \cdot h \cdot w)}$, which are seen as sequences of $n_{t_a} \cdot n_d$ and $n_{t_v} \cdot n_h \cdot n_w$ tokens of dimension $t_a \cdot d$ and $t_v \cdot h \cdot w$, respectively. 

The use of discrete audio and visual tokens in VQ-MAE-AV has several motivations. Firstly, by dividing the audiovisual data into spatio-spectro-temporal patches, the method could learn to relate audio tokens to visual tokens. For example, the audio tokens are expected to correlate strongly with the visual tokens corresponding to the mouth area \citep{arnela2016influence, sadok2024multimodal}. The proposed method can potentially learn a representation that captures shared and distinctive information between the two modalities by exploiting their complementarity. Additionally, the use of discrete tokens can reduce the computational cost of the method as it involves working with a reduced representation of the data, which allows us to increase the number of tokens (i.e., manipulate longer audiovisual speech sequences) without exploding in the number of trainable parameters compared to the multimodal MAE in the literature~\citep{bachmann2022multimae}. 

\subsection{Masking}
\label{subsec:masking}
 
The masking strategy employed to train the MAE will impact the performance of downstream tasks \cite{he2022masked}. In the original MAE, the masked tokens are chosen randomly given a target masking ratio, typically 75\%. The MAE is then trained to reconstruct the masked tokens from the visible ones, and in doing so it learns a representation that can be used for downstream tasks. In a multimodal scenario, the straightforward approach would be to apply this same masking strategy independently on each modality. However, it has been shown that it is more effective to implement a coupled masking strategy between the modalities \citep{bachmann2022multimae}. Therefore, to the train the VQ-MAE-AV model, the masking ratio for the audio and visual modalities is drawn randomly according to a uniform distribution on the 1-simplex. This means that if $p \times 100~\%$ of the tokens are masked in one modality, then $(1-p) \times 100~\%$ of the tokens are masked in the other modality, where $p$ is distributed uniformly between 0 and 1.
This strategy allows the reconstruction of missing information from one modality by relying on the other.

\begin{figure*}[h!]
    \centering
    \includegraphics[width=0.9\textwidth]{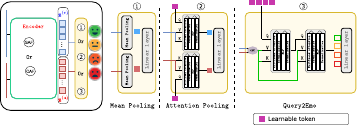}
    \caption{Overview of the three emotion recognition models trained on top of the VQ-MAE-AV encoder.}
    \label{fig:fine_tuing}
\end{figure*}

\subsection{Continuous embedding vectors}
\label{subsec:embedding}

The discrete tokens correspond to the indices obtained through the quantization step of the pretrained VQ-VAE encoder. Before being input into the VQ-MAE-AV encoder, these discrete tokens are replaced with trainable continuous embedding vectors taken from an audio codebook in $\mathbb{R}^{k_a \times e_a}$ and from a visual codebook in $\mathbb{R}^{k_v \times e_v}$, where $k_{a/v}$ is the number of codes in the codebook and $e_{a/v}$ is the dimension of each code. This is simply achieved by replacing the indices of a discrete token with the corresponding vectors of dimension $e_{a/v}$ in the codebook. 
After this embedding process, the sequences of discrete tokens $\xaq \in \mathbb{N}^{(n_{t_a} \cdot n_d) \times (t_a \cdot d)}$ and $\xvq$ in $\mathbb{N}^{(n_{t_v} \cdot n_h \cdot n_w) \times (t_v \cdot h \cdot w)}$ are transformed into sequences of continuous tokens $\xac \in \mathbb{R}^{(n_{t_a} \cdot n_d) \times ({t_a} \cdot d \cdot e_a)}$ and $\xvc \in \mathbb{R}^{(n_{t_v} \cdot n_h \cdot n_w) \times ({t_v} \cdot h \cdot w \cdot e_v)}$. The dimensions $e_{a}$ and $e_{v}$ of the audio and visual codes are chosen such that the continuous tokens in both modalities have the same dimension $E = {t_a} \cdot d \cdot e_a = {t_v} \cdot h \cdot w \cdot e_v $, which is necessary for concatenation in the VQ-MAE-AV encoder (see the next subsection). Therefore, we have $\xac \in \mathbb{R}^{(n_{t_a} \cdot n_d) \times E}$ and $\xvc \in \mathbb{R}^{(n_{t_v} \cdot n_h \cdot n_w) \times E}$.

\subsection{VQ-MAE-AV encoder and decoder}
\label{subsec:encoder_decoder}

\subsubsection{Attention Block}
\label{subsubsec:attention_block}

The VQ-MAE-AV encoder and decoder are built with multi-head Attention blocks similar to those used in the Vision Transformer (ViT) \citep{dosovitskiy2020image}. Each block comprises a multi-head attention layer, normalization layers, and a Multi-layer Perceptron (MLP). These layers are interconnected with residual connections, as depicted in Fig. \ref{fig:Overview}, and they will be used to capture the inter and intra-relationships between audio and visual tokens. This attention block is inspired by the attention layer in the original transformer \citep{vaswani2017attention}.
To simplify the reading afterward, we denote the attention block by $\texttt{Attention}(Q, V, K)$, where $Q, V, K$ are the query, value, and key, respectively. The self-attention mechanism uses the same input vector for the query, key, and value vectors. In the case of cross-attention, the query and key are different to enable attention across multiple modalities \textit{or} inputs.

\subsubsection{Encoders}\label{subsubsec:encoder} 

We propose two fusion strategies of the audio and visual speech data, resulting in two architectures for the VQ-MAE-AV encoder. The first fusion strategy is called \emph{self-attention fusion}. As represented in Fig.~\ref{fig:Overview}, this fusion consists of a concatenation of token sequences for the two modalities, followed by $L$ self-attention blocks. This concatenation operates on the first dimension of the variables, i.e., we obtain a sequence of $(n_{t_a} \cdot n_d) + (n_{t_v} \cdot n_h \cdot n_w)$ tokens after concatenation.

The second fusion strategy is called $\emph{cross-attention fusion}$. As shown in Fig.~\ref{fig:Overview}, the sequences of audio and visual tokens are used separately as the queries of two separate attention blocks, which share the same keys and values corresponding to the concatenation of the modalities. These two cross-attention blocks are then followed by a stack of $L$ self-attention blocks. 

For both fusion strategies, the encoder outputs one sequence of tokens for each modality, denoted by $\za$ and $\zv$.

\subsubsection{Global tokens}\label{subsubsec:global_token} 

In addition to the token-wise representations $\za$ and $\zv$, we learn two sequence-wise global tokens, denoted by $\zaG \in \mathbb{R}^{1 \times E}$ and $\zvG \in \mathbb{R}^{1 \times E}$, which can be thought of as similar to [CLS] tokens \citep{he2022masked}. These global modality-specific tokens are introduced to aggregate the spectro-temporal and spatio-temporal information in the two modalities, which can be useful for downstream tasks involving predictions at the sequence level, such as audiovisual SER. The global tokens are computed with attention pooling as follows:
\begin{align}
    \zaG &=\texttt{Attention}\left(Q_{(a)}, V_{(a)}, K_{(a)} \right);\\
    \zvG &=\texttt{Attention}\left(Q_{(v)}, V_{(v)}, K_{(v)} \right),
\end{align}
where $Q_{(a)} \in \mathbb{R}^{1 \times E}$ and $Q_{(v)} \in \mathbb{R}^{1 \times E}$ represent respectively trainable audio and visual tokens, as proposed in \cite{touvron2021augmenting}, $V_{(a)} = K_{(a)} = \za$, and $V_{(v)} = K_{(v)} = \zv$.

\subsubsection{Decoders}\label{subsubsec:decoder} 
The token-wise representation obtained from the encoder is combined with mask tokens and fed to the VQ-MAE-AV decoder along with additional position embeddings, as denoted by $\tildza$ and $\tildzv$ and illustrated in Fig.~\ref{fig:Overview}. The mask tokens actually correspond to one single trainable vector as proposed in the original MAE \citep{he2022masked}. Similarly as for the encoder, the audio and visual inputs of the VQ-MAE-AV decoder can be fused using either self-attention fusion or cross-attention fusion.

The number of attention blocks $L'$ in the decoder is chosen to be lower compared to that of the encoder ($L' < L$).

A linear layer is added at the end of the decoder, which maps to the size of the VQ-VAE codebooks. The output of this linear layer corresponds to the logits of the discrete tokens. After applying a \textit{argmax} operation, we obtain reconstructions $\xaqrec$ and $\xvqrec$ of the indices $\xaq$ and $\xvq$ that were provided by the VQ-VAE-A and VQ-VAE-V encoders, respectively. 

\subsection{Loss functions}
\label{subsec:loss_function}

The VQ-MAE-AV model is trained in a self-supervised manner to minimize a generative loss $\mathcal{L}_{gen}$ between the reconstructed and original tokens and a contrastive loss $\mathcal{L}_{NCE}$ between the audio and visual global tokens.

Due to their discrete nature, the cross-entropy (CE) is naturally used to measure the reconstruction quality of the audiovisual masked tokens:
\begin{align}
    \label{eq:generative-loss}
    \mathcal{L}_{gen}&\Big(\xaq, \xaqrec, \xvq, \xvqrec,\Omega^{(a)}, \Omega^{(v)} \Big) = \nonumber \\ &\texttt{CE}\Biggl(\xaq \Big(\Omega^{(a)}\Big),~\xaqrec \Big(\Omega^{(a)}\Big)\Biggl) \, + \, \texttt{CE}\Biggl(\xvq\Big(\Omega^{(v)} \Big),~\xvqrec\Big(\Omega^{(v)}\Big)\Biggl),
\end{align}
where $\mathbf{x}\left(\Omega^{(\cdot)}\right)$ denotes the set of masked tokens in $\mathbf{x}$. As can be seen, another benefit of manipulating discrete representations for multimodal inputs is the homogeneity of the losses, which does not require balancing the losses between the two modalities.

Building upon the approaches presented in \cite{alayrac2020self, akbari2021vatt}, the global tokens are learned using noise contrastive estimation. This approach enhances the alignment of audiovisual speech pairs by grouping together embeddings that belong to the same time sequence and separating them from those that do not correspond to the same sequence. It involves minimizing the loss function $\mathcal{L}_{NCE}(\zaG, \zvG)$, which is defined by:
\begin{align}
    \label{eq:nce}
    &\mathcal{L}_{NCE}(\mathbf{u}, \mathbf{v}) = -\log \left(\frac{\exp \Bigl( \mathbf{u}^\top \mathbf{v}/\tau \Bigl)}{\exp \Bigl( \mathbf{u}^\top \mathbf{v}/\tau \Bigl) + \sum\limits_{(\mathbf{u}',~\mathbf{v}') \in \mathcal{N}} \exp \Bigl( \mathbf{u}'^\top \mathbf{v}'/\tau \Bigl)}\right).
\end{align}

To form positive pairs $(\zaG,~\zvG)$ for both audio and visual modalities, we select corresponding streams from the same temporal location in the video. Conversely, negative pairs $(\zaGN,~\zvGN)$ are formed by selecting \textit{non}-corresponding streams drawn from a set $\mathcal{N}$ of different temporal locations for each batch. The sensitivity of the NCE loss in distinguishing between positive and negative pairs is regulated by a temperature parameter $\tau \in \mathbb{R}$.

The overall loss function used to train VQ-MAE-AV is simply the sum of the generative and contrastive loss functions in \eqref{eq:generative-loss} and \eqref{eq:nce}.

\subsection{Emotion recognition}
\label{subsec:finetuning}

After self-supervised learning on a large-scale unlabeled dataset, the VQ-MAE-AV model is used for emotion recognition. The task is to predict the emotion class of an input audiovisual speech sequence. To address this task, we need to introduce a small emotion recognition model for predicting the emotion category from the audiovisual speech representation provided by the pre-trained VQ-MAE-AV encoder. In this work, we propose and investigate three different emotion recognition models, which are described in the remainder of this section and illustrated in Fig.~\ref{fig:fine_tuing}. All emotion recognition models are trained for a supervised classification task using the asymmetric loss of \citet{ridnik2021asymmetric}. Along with the training of the emotion recognition model, we also fine-tune the VQ-MAE-AV encoder.

\subsubsection{Emotion recognition model based on attention pooling}

The first approach to performing emotion recognition from the proposed VQ-MAE-AV model is to use the sequence-wise global tokens $\zaG$ and $\zvG$, which were computed using attention pooling of the token-wise representations $\za$ and $\zv$, as described in Section~\ref{subsubsec:global_token}. In this approach, the two global tokens are concatenated and passed through a single linear layer followed by a softmax operation to predict the emotion class probabilities. 

\subsubsection{Emotion recognition model based on mean pooling}

The second approach for emotion recognition replaces the previous attention pooling strategy with a more naive one, where the token-wise representations $\za$ and $\zv$ are aggregated temporally using a simple parameter-free mean pooling operation. The resulting vectors are then concatenated and passed through a single linear layer, as before.

\subsubsection{Emotion recognition model based on Query2Emo}

Finally, we propose a third emotion recognition approach, referred to as Query2Emo and inspired by \citep{liu2021query2label}. As illustrated in Fig.~\ref{fig:fine_tuing}, Query2Emo involves cross-attention between all audio and visual tokens (concatenation of $\za,~\zv$) as key and value, and the emotion classes represented by trainable embeddings as the query:
\begin{align}
\mathbf{w}_{emo} = \texttt{Attention}\left(Q_{emo}, \mathbf{z}, \mathbf{z} \right),
\end{align}
where $\mathbf{z}$ denotes the concatenation of the two sequences $\za$ and $\zv$, and $Q_{emo} \in \mathbb{R}^{K_{emo} \times E}$ corresponds to $K_{emo}$ trainable tokens of dimension $E$, with $K_{emo}$ the number of emotion classes. This cross-attention aims to learn the relationship between the emotion representation $Q_{emo}$ and the audiovisual tokens for the SER task. Query2Emo consists of two attention blocks as illustrated in Fig.~\ref{fig:fine_tuing}. The model provides at the output $K_{emo}$ vectors of dimension $E$ (i.e., $\mathbf{w}_{emo} \in \mathbb{R}^{K_{emo} \times E})$, which are then concatenated and passed through a single linear layer, as for the two previous emotion recognition models.


\section{Experiments}
\label{sec:Experiments}

This section starts by presenting the experimental setup, including the datasets and preprocessing, the model architecture, and the training configuration. Then, we present a first experiment that aims to measure the reconstruction quality of the VQ-MAE-AV model when applied to masked audiovisual speech data. We then evaluate the performance of the proposed approach for emotion recognition using four audiovisual speech datasets and comparing with several state-of-the-art methods. Finally, we present the results of an ablation study that measures the impact of various hyperparameters and architecture choices on the performance of our method. 

\subsection{Experimental setup}   
\label{sec:pretrain_setting}

\subsubsection{Datasets and preprocessing}
\paragraph{Dataset for self-supervised training} To pre-train VQ-MAE-AV, we use the VoxCeleb2 dataset \citep{chung2018voxceleb2}, which offers a broad range of audiovisual speech data from open-source media, with each video featuring a single speaker.   We restricted our dataset use to a subset of around 1000 hours of audiovisual speech, encompassing 2170 different speakers. The test set includes about 100 hours of audiovisual speech data with 117 different speakers.Wild2

\paragraph{Data pre-processing} The VQ-VAE-A and VQ-VAE-V models are trained on the VoxCeleb2 dataset. The former is trained on short-time Fourier transform (STFT) power spectrograms ($\xa$), while the latter is trained on sequences of RGB images ($\xv$) captured at 25~fps, cropped and resized to a resolution of $96 \times 96$, and aligned using Face-Alignment \citep{bulat2017far}. To compute the STFT, a Hann window of 64 ms (1024 samples at 16 kHz) and a $68$\% overlap are used, resulting in a sample rate of 50 Hz, which is twice the sample rate of the visual modality. This leads to sequences of $D=513$ Fourier coefficients.

\paragraph{Emotional audiovisual speech datasets} We fine-tune and evaluate the proposed approach for audiovisual SER in both controlled and in-the-wild conditions, using four datasets that are presented below.
\begin{itemize}
    \item RAVDESS \citep{livingstone2018ryerson} is a dataset that consists of 1,440 videos recorded by 24 English-speaking actors in controlled conditions. It is labeled with eight emotions: neutral, calm, happy, sad, angry, fearful, disgust, surprised. 
    \item CREMA-D \citep{cao2014crema} is a dataset that consists of 7,442 videos recorded by 91 English-speaking actors in controlled conditions. Actors spoke from a selection of 12 sentences, using one of six emotions (anger, disgust, fear, happy, neutral, and sad) with four different intensities (low, medium, high, and unspecified).
    \item DFEW \citep{jiang2020dfew} is a dataset containing 16,372 video clips of English-speaking subjects, annotated with seven emotions: neutral, happy, sad, surprise, fear, anger, and disgust. The clips are extracted from diverse movie scenes, ensuring a wide range of variations in expressions, lighting, and occlusions, providing a challenging benchmark for emotion recognition in the wild.
    \item Aff-Wild2 \citep{kollias2018aff}: Aff-Wild2 comprises 564 videos extracted from Youtube (around 2.8M frames), including 554 different English-speaking subjects, and annotated with seven emotions: neutral, anger, disgust, fear, happiness, sadness, surprise. Similar to DFEW, Aff-Wild2 captures spontaneous expressions in uncontrolled environments, including diverse recording conditions, making it suitable for emotion recognition in real-world scenarios.
\end{itemize}
We selected these datasets as they are commonly used for the emotion recognition task, and the raw audiovisual speech data is accessible. We performed \emph{6}-fold and \emph{10}-fold cross-validation for the RAVDESS and CREAM-D datasets to ensure a fair comparison with previous works. For DFEW and Aff-Wild2, the train, validation, and test splits are already defined. For all datasets, the speaker identities are different in the training and evaluation sets, allowing us to evaluate SER in a speaker-independent setting.

\begin{figure*}[t!]
     \begin{subfigure}[b]{0.48\textwidth}
         \centering
         \includegraphics[width=\textwidth]{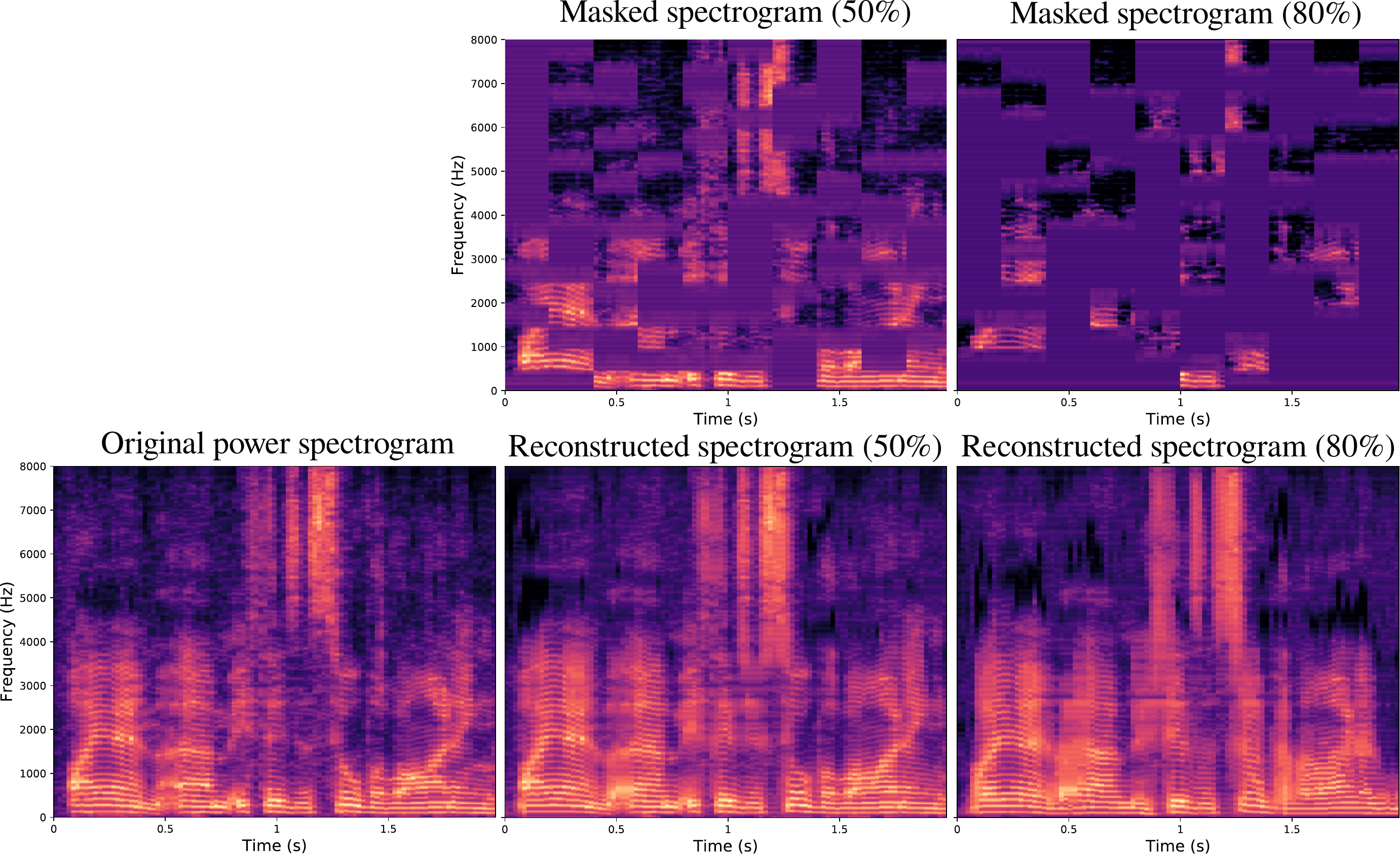}
        \caption{Qualitative reconstruction results for the audio modality. The spectrogram highlighted in the red box represents the original spectrogram. The two spectrograms on the top right represent the spectrograms masked at 50\% and 80\%, respectively. The reconstructions using VQ-MAE-AV can be seen directly below these masked spectrograms.}
        \label{fig:qualitative-audio}
     \end{subfigure}
     \hfill
     \begin{subfigure}[b]{0.48\textwidth}
         \centering
        \includegraphics[width=\textwidth]{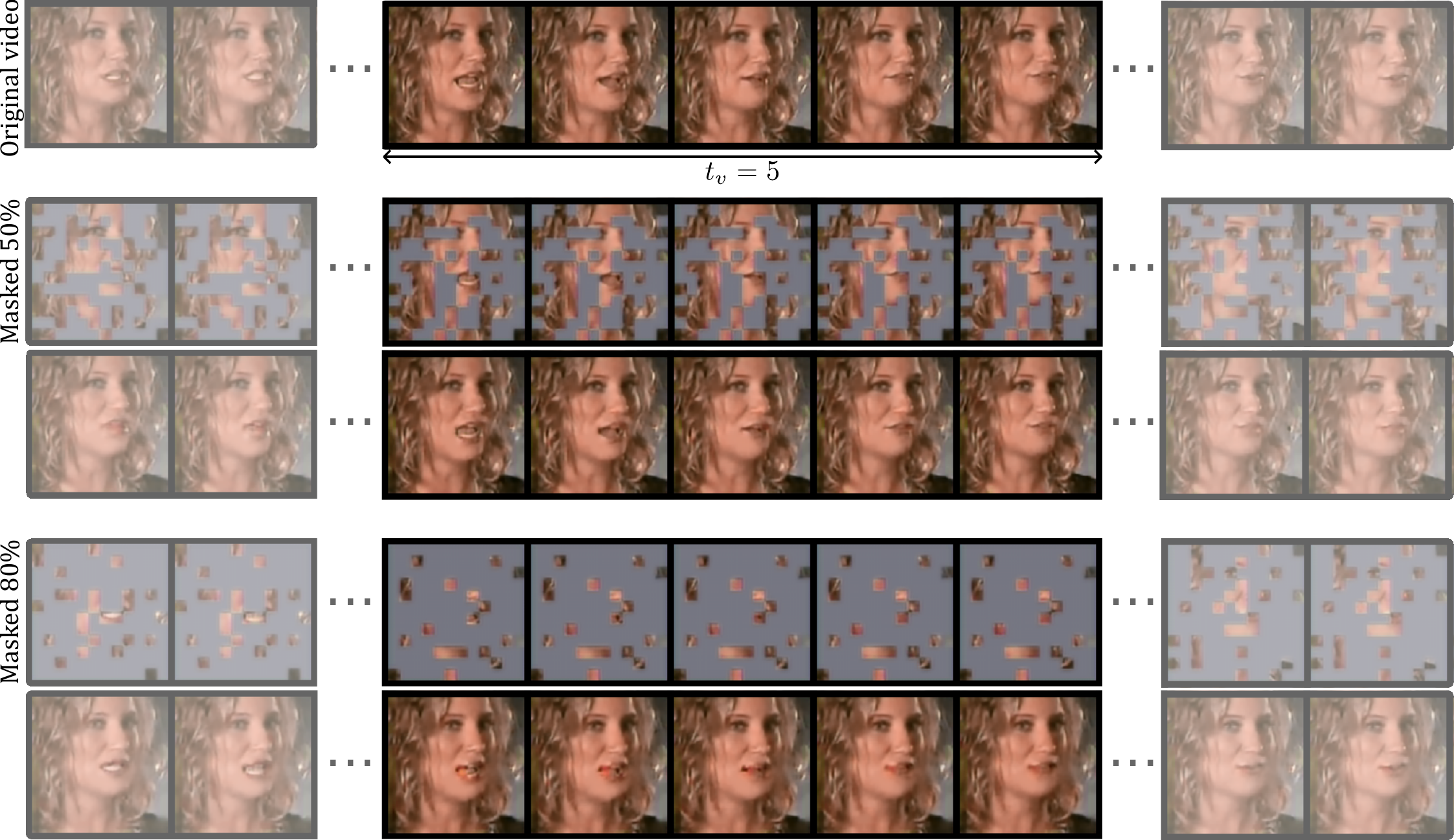}
        \caption{Qualitative reconstruction results for the visual modality. The first sequence shows the original video, followed by the next two sequences representing the masked video with a ratio of 50\% and its reconstruction using VQ-MAE-AV. The last two sequences represent the masked video with a ratio of 80\% and its reconstruction using VQ-MAE-AV.}
        \label{fig:qualitative-visual}
     \end{subfigure}
        \caption{Quantitative results of the audio reconstruction (a) and visual reconstruction (b) using the VQ-MAE-AV model.}
        \label{fig:qualitative}
\end{figure*}

\begin{figure*}[t!]
     \begin{subfigure}[b]{0.48\textwidth}
         \centering
         \includegraphics[width=\textwidth]{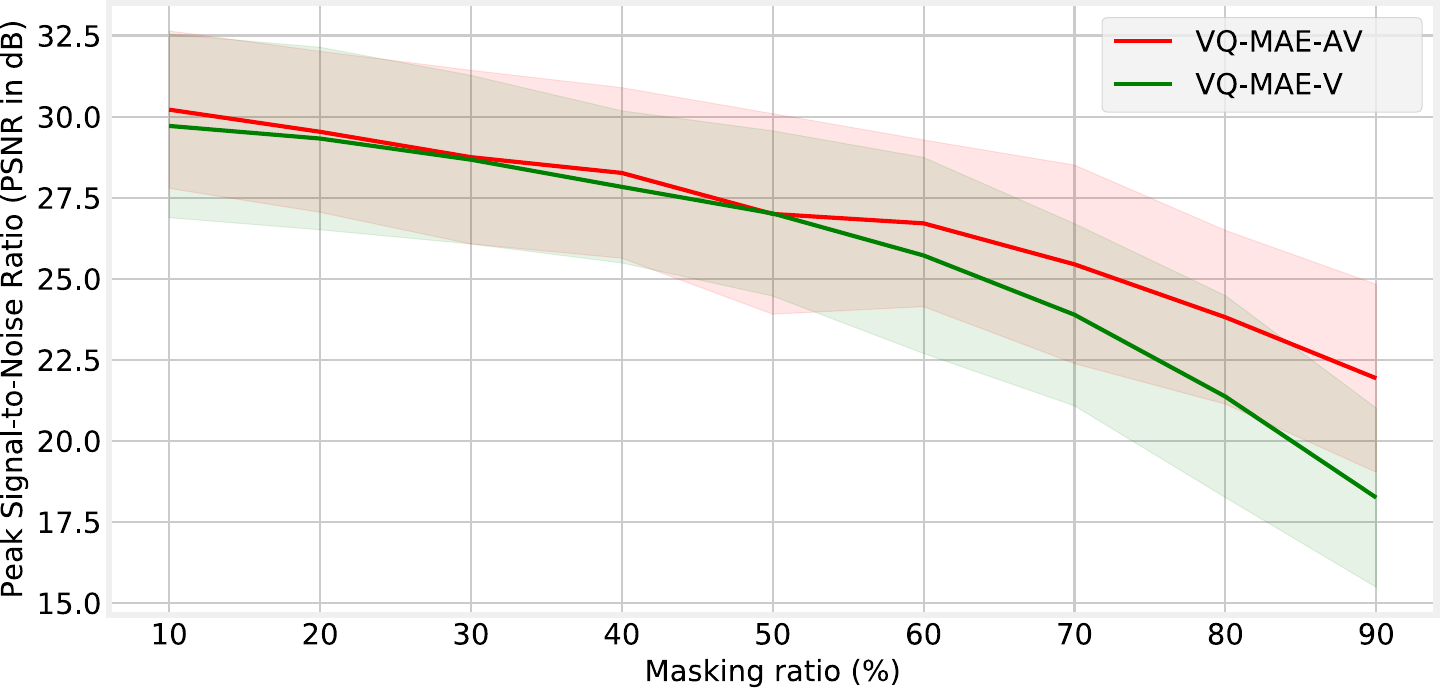}
         \caption{Peak Signal-to-Noise Ratio (PSNR in dB) on the y-axis as a function of masking ratio (\%) on the x-axis.}
         \label{fig:unmasking-visual}
     \end{subfigure}
     \hfill
     \begin{subfigure}[b]{0.48\textwidth}
         \centering
         \includegraphics[width=\textwidth]{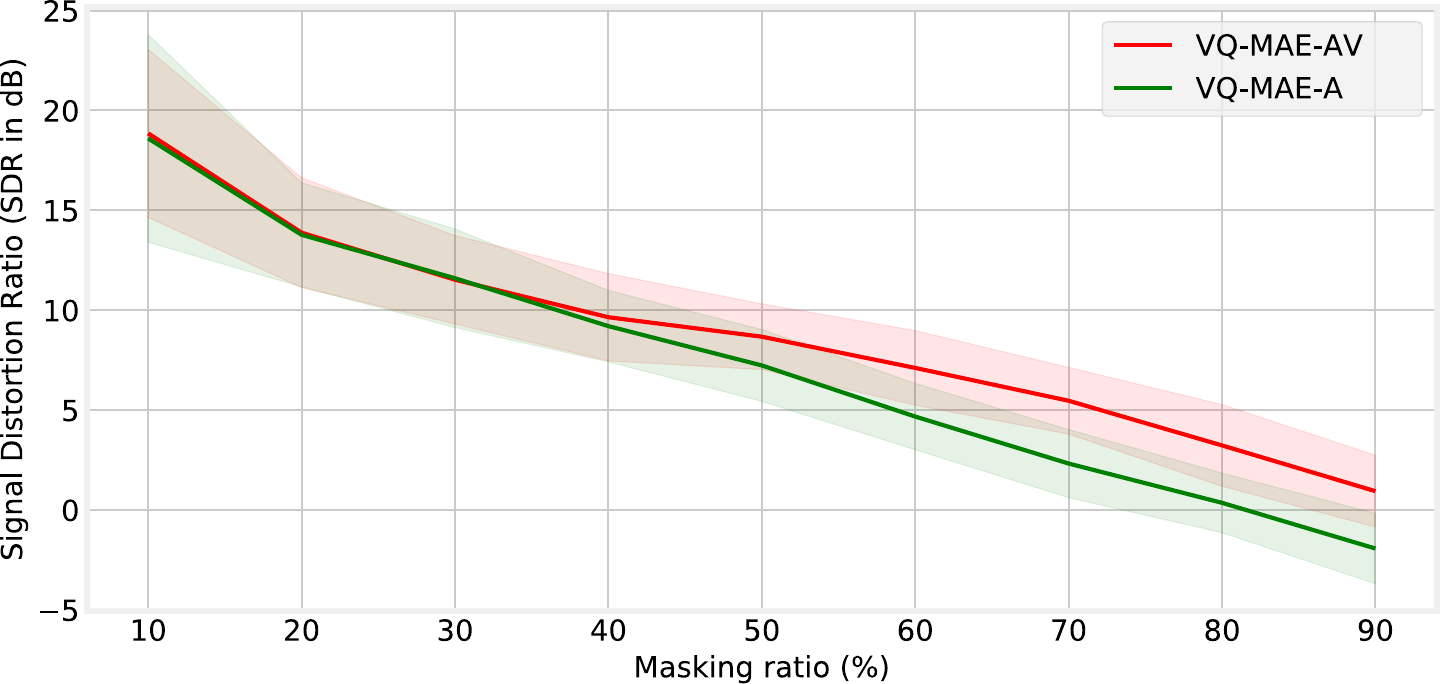}
         \caption{Signal-to-Distortion Ratio (SDR in dB) on the y-axis as a function of masking ratio (\%) on the x-axis.}
         \label{fig:unmasking-audio}
     \end{subfigure}
        \caption{Quantitative results of the visual reconstruction (a) and audio reconstruction (b). The line represents the mean value across the test dataset, while the shaded area depicts the standard deviation.}
        \label{fig:unmasking}
\end{figure*}

\subsubsection{Model architecture}
\label{sec:architecture}
\paragraph{VQ-VAE architectures} The VQ-VAE-A (respectively VQ-VAE-V) architecture is symmetrical concerning the encoder and the decoder, with three 1D (respectively 2D) convolution for the encoder or transposed convolution for the decoder layers and a residual convolution layer. The VQ-VAE models process each frame independently with no time dependency. For each speech power spectrogram frame of size $D=513$, the VQ-VAE-A encoder compresses it into a discrete latent vector (a column of $\xaq$) of size $D'=64$. For each image frame of size $(H=96, W=96, C=3)$, the VQ-VAE-V encoder compresses it into a discrete latent representation of size $(H'=24, W'=24)$. 
The VQ-VAE-A and the VQ-VAE-V codebooks contain, respectively, $k_a=256$ and $k_v=512$ codes of dimension $e_a=8$ and $e_v=4$. Such a low dimension is chosen to increase the use of the different codes in the codebook \citep{yuvector}. 

\paragraph{VQ-MAE-AV architectures} 

The VQ-MAE-AV model employs an encoder comprising $L$ attention blocks, where $L$ is set to 12 unless otherwise specified, and a decoder featuring 4 attention blocks. Each self-attention layer of a block is divided into 4 heads. By default, the parameters of the discrete audio and visual tokens ($d$, $h$, and $w$) are set to 4. We set $t_a = 10$ and $t_v = 5$ because the sampling rate of $\xa$ is twice the sampling rate of $\xv$.
In the ablation study (Section~\ref{sec:ablation}), we will explore all possible combinations of the VQ-MAE-AV encoder and decoder, according to the two fusion strategies (\emph{self-attention fusion} and $\emph{cross-attention fusion}$). We will also evaluate the impact of the pooling strategy and contrastive learning. 

\subsubsection{Training and fine-tuning details}
\paragraph{Self-supervised training details} The VQ-MAE-AV is trained using the AdamW optimizer \citep{loshchilov2017decoupled} with a cosine scheduler to adjust the learning rate, with a 100-epoch warm-up period. The parameters of the optimizer, similar to \cite{he2022masked}, are $\beta_2=0.9$, $\beta_2=0.95$, and \texttt{weight\_decay}$=0.05$. The base learning rate follows the linear scaling rule \citep{goyal2017accurate} $lr = (\texttt{base\_lr}=1e-3) \times (\texttt{batchsize}=128)/256$. We distributed the pre-training of VQ-MAE-AV on 4 NVIDIA HGX A100. The training lasted for 160 epochs, and each epoch took approximately 15 minutes.

\paragraph{Fine-tuning details} For the fine-tuning process, we also use the AdamW optimizer \citep{loshchilov2017decoupled} with a cosine scheduler to adjust the learning rate and with a 40-epoch warm-up period. The parameters of the optimizer are the same as those used for the pre-training. The base learning rate is \texttt{1e-4}.

\subsection{Audiovisual speech reconstruction quality}
\label{sec:rec-quality}

In this experiment, we evaluate the reconstruction quality of VQ-MAE-AV when applied to masked audiovisual speech data. The model is fed with a sequence of tokens, some of which have been masked, and it is used to predict the masked tokens, i.e. to reconstruct the complete audiovisual speech sequence from partially observed tokens. We are interested in studying the reconstruction performance for different masking ratios. Qualitative results are presented in Fig.~\ref{fig:qualitative-audio} and Fig.~\ref{fig:qualitative-visual} for two masking ratios, 50\% and 80\%. Additional qualitative results are available online (see the webpage address at the end of Section~\ref{sec:intro}).

In this experiment, we compare VQ-MAE-AV to VQ-MAE-A (A for audio) and VQ-MAE-V (V for visual), which are the unimodal versions of VQ-MAE-AV. The average quality performance for the speech and visual modalities is evaluated using the VoxCeleb2 test set. The Peak Signal-to-Noise Ratio (PSNR in dB) is used to assess the quality of the resynthesized visual data, and the Signal-to-Distortion Ratio (SDR in dB) is used to assess the quality of the resynthesized audio data.

Fig.~\ref{fig:unmasking-visual} and Fig.\ref{fig:unmasking-audio} present the PSNR and SDR curves, respectively, for the reconstruction quality of the visual and audio modalities as a function of the masking ratio. Notably, VQ-MAE-AV outperforms VQ-MAE-V for masking ratios greater than $50$\%. At a masking ratio of $90$\%, VQ-MAE-AV achieves a significant $3.68$~dB gain in PSNR over VQ-MAE-V. For the audio modality, VQ-MAE-AV outperforms VQ-MAE-A for masking ratios greater than $40$\% and records a gain of $2.87$~dB in SDR at $90$\% of masking. In summary, this experiment highlights the effectiveness of leveraging multimodality to improve reconstruction quality.

\subsection{Audiovisual emotion recognition}
\label{sec:emotion_recognition}

\begin{figure*}[t]
\begin{minipage}{1.0\textwidth}
\centering
\vspace{0.2cm}
\captionof{table}{Audiovisual emotion recognition results in terms of accuracy (\%) and F1 score (\%) on the RAVDESS and CREMA-D datasets}. The best scores are in bold, and the second-best scores are underlined.
\label{tab:emo_recog_results}
\begin{tabular}{cccccc}
\hline \hline
\multicolumn{3}{c}{RAVDESS} & \multicolumn{3}{c}{CREMA-D} \\
\cmidrule(lr){1-3} \cmidrule(lr){4-6}
\multicolumn{1}{c}{Method} & \multicolumn{1}{c}{Accuracy} & \multicolumn{1}{c}{F1 score} & \multicolumn{1}{c}{Method}  & \multicolumn{1}{c}{Accuracy} & \multicolumn{1}{c}{F1 score}  \\
\cmidrule(lr){1-3} \cmidrule(lr){4-6} 
AV-LSTM \citep{ghaleb2019multimodal}  &  65.80 & - & AV-LSTM \citep{ghaleb2019multimodal}  & 72.90 & -  \\
MuLT \citep{tsai2019multimodal}  &  76.60 & 77.30 & MATER \citep{ghaleb2020multimodal}  & 67.20 & -  \\
CFN-SR \citep{fu2021cross}  &  75.76 & - & AV-Gating  \citep{ghaleb2019multimodal} & 74.00 & - \\
MATER \citep{ghaleb2020multimodal}  &  76.30 & - & MulT Base \citep{tran2022pre}  &  68.87 & -\\
AVT \citep{chumachenko2022self}  &  79.20 & 78.20 & MulT Large \citep{tran2022pre}  & 70.22 & - \\
MDVAE \citep{sadok2024multimodal}  &   \uline{79.30} & \uline{80.70} &  RAVER \citep{goncalves2022robust}  & \uline{77.30} & - \\
\cmidrule(lr){1-3} \cmidrule(lr){4-6}
VQ-MAE-AV (ours) &  \textbf{84.80} & \textbf{84.50} & VQ-MAE-AV (ours)& \textbf{80.40} & \textbf{80.00} \\
[.2cm]
\hline \hline
\end{tabular}
\end{minipage}
\end{figure*}

\begin{figure*}[t]
\begin{minipage}{1.0\textwidth}
\centering
\vspace{0.2cm}
\captionof{table}{Accuracy (\%) and F1 score (\%) results on DFEW and Aff-Wild2. The best scores are in bold, and the second-best scores are underlined.}
\label{tab:emo_recog_dfew_affwild}
\begin{tabular}{cccccc}
\hline \hline
\multicolumn{3}{c}{DFEW} & \multicolumn{3}{c}{Aff-Wild2} \\
\cmidrule(lr){1-3} \cmidrule(lr){4-6}
\multicolumn{1}{c}{Method} & \multicolumn{1}{c}{Accuracy} & \multicolumn{1}{c}{F1 score} & \multicolumn{1}{c}{Method}  & \multicolumn{1}{c}{Accuracy} & \multicolumn{1}{c}{F1 score}  \\
\cmidrule(lr){1-3} \cmidrule(lr){4-6} 
C3D+LSTM \citep{zhang2023transformer}   &  65.17 & - & CT-VIPL \citep{liu2020emotion}  & 64.00 & 33.00  \\
ResNet-18+LSTM \citep{zhang2023transformer}  &  64.32 & - & Multi-modal ER \citep{jin2021multi}  & 63.00 & 40.20  \\
T-MEP \citep{zhang2023transformer} &  \uline{68.85} & - & CNN-RNN \citep{antoniadis2021audiovisual}  & \uline{66.80} & \uline{55.50} \\
\cmidrule(lr){1-3} \cmidrule(lr){4-6}
VQ-MAE-AV (ours) &  \textbf{70.30} & \textbf{69.80} & VQ-MAE-AV (ours) & \textbf{69.70} & \textbf{68.90} \\
[.2cm]
\hline \hline
\end{tabular}
\end{minipage}
\end{figure*}

In this section, we evaluate the emotion recognition performance of the proposed VQ-MAE-AV model on 4 different datasets, including 2 recorded in the wild, and we compare the model's performance with that of 15 state-of-the-art methods. The performance is measured in terms of accuracy and F1 score. For this experimental comparison, we chose the best configuration of the VQ-MAE-AV model, which uses the \emph{cross-attention fusion} strategy for the encoder and decoder, both generative and contrastive loss functions for self-supervised pre-training on VoxCeleb2, and the Query2Emo strategy for supervised fine-tuning. Other configurations will be investigated in the ablation study of Section~\ref{sec:ablation}.

\subsubsection{Results in controlled conditions}

We start by discussing the emotion recognition results on RAVDESS and CREMA-D, the two datasets recorded in controlled lab conditions.
The experimental comparison includes LSTM-based methods \citep{ghaleb2019multimodal} and Transformer-based methods \citep{tsai2019multimodal, chumachenko2022self, goncalves2022robust}. MATER \citep{ghaleb2020multimodal} uses distinct Transformer architectures for each modality, merging the embeddings via mean pooling. CFN-SR \citep{fu2021cross} introduces a cross-modality learning approach, employing self-attention and residual structures to model inter- and intra-modality interactions. We also consider the AVT method \citep{chumachenko2022self}, which uses self-attention fusion and modality dropout to address the challenge of one modality being absent. Additionally, RAVER \citep{goncalves2022robust} addresses challenges related to modality alignment, temporal information capture, and handling missing features.

The experimental comparison also includes the multimodal dynamical VAE (MDVAE) \citep{sadok2024multimodal}, which employs an unsupervised hierarchical latent representation to segregate static from dynamic information and modality-common from modality-specific information. Additionally, we compare with MulT \citep{tran2022pre}, a self-supervised approach using the Transformer architecture and pre-trained on the VoxCeleb dataset with a pretext task of reconstructing masked frames (with a masking ratio of 15\%).

As can be seen in Table~\ref{tab:emo_recog_results}, the VQ-MAE-AV model outperforms the most recent methods overall. 
VQ-MAE-AV achieves $9.04$\%, $5.60$\% and $5.50$\% better accuracy than the CFN-SR \citep{fu2021cross}, AVT \citep{chumachenko2022self} and MDVAE \citep{sadok2024multimodal}  methods for the RAVDESS dataset, respectively.
Regarding the CREMA-D dataset, VQ-MAE-AV achieves $7.50$\%, $6.40$\%, and $3.10$\% better accuracy than the AV-LSTM \citep{ghaleb2019multimodal}, AV-Gating \citep{ghaleb2019multimodal}, and RAVER \citep{goncalves2022robust}. 

\subsubsection{Results on in-the-wild datasets}

We now discuss the emotion recognition results of the proposed approach in more realistic recording conditions, using the two in-the-wild datasets DFEW and Aff-Wild2. The experimental comparison includes methods based on combinations of convolutional and LSTM networks \citep{liu2020emotion, jin2021multi, zhang2023transformer, antoniadis2021audiovisual} and a Transformer-based method \citep{zhang2023transformer}.

As can be seen in Table~\ref{tab:emo_recog_dfew_affwild}, the VQ-MAE-AV model with Query2Emo achieves superior performance compared to recent methods on the DFEW dataset. Specifically, VQ-MAE-AV outperforms the Transformer-based method T-MEP \citep{zhang2023transformer} by $1.45\%$ in accuracy. On the Aff-Wild2 dataset, VQ-MAE-AV achieves a $2.9\%$ improvement in accuracy and a $13.4\%$ boost in F1 score compared to the CNN-RNN model of \citet{antoniadis2021audiovisual}. These results confirm that the VQ-MAE-AV model also performs well on real audiovisual speech data recorded in uncontrolled environments.

Overall, the experimental results demonstrate the effectiveness of the proposed audiovisual self-supervised representation learning technique for SER. This indicates that VQ-MAE-AV learns audiovisual representations that are effectively transferable to the emotion recognition task, resulting in improved performance compared to state-of-the-art methods.

\begin{table*}[t]
    \centering
    \begin{minipage}{.49\linewidth}
      \centering
        \caption{Performance of the baseline VQ-MAE-AV model (first row, in blue), compared to its performance when the encoder is frozen during supervised training on RAVDESS (second row) or when self-supervised pre-training on VoxCeleb2 is discarded (third row).}
        \label{tab:finetuning}
            \begin{tabular}{ccc}
                \hline
                Pre-training  & Frozen encoder  & Accuracy \small{(\%)} \\ \hline
                                 \color{blue}\checkmark             & \color{blue}\ding{55}             & \color{blue}\textbf{81.5} \\
                \checkmark             & \checkmark            & 70.5 \\
                \ding{55}               & \ding{55}             & 29.6 
            \end{tabular}
    \end{minipage}
    \hfill \vspace{.5cm}
    \begin{minipage}{.49\linewidth}
      \centering
        \caption{Performance, number of parameters (in millions), and floating-point operations (FLOPs, in billions) of the baseline VQ-MAE-AV model (second row, in blue), compared to variations where the number $L$ of attention blocks in the encoder is modified (other rows).}
        \label{tab:depth}
            \resizebox{1.0\linewidth}{!}{ 
            \begin{tabular}{ccccccc}
                \hline
                Encoder depth          & Param. (M) & FLOPs (G) & runtime (ms) & Acc. \small{(\%)} & F1 score \small{(\%)}\\ \hline
                $L=6$       &     8.5  &   5.9 & 0.37  & 75.7  & 76.0\\
                \color{blue}$L=12$     &     \color{blue}13.5  &   \color{blue}9.4 & \color{blue}0.67   & \color{blue}81.5 & \color{blue}80.1 \\
                $L=16$       &     16.8   &   11.7 & 0.86 & \textbf{82.4}  & \textbf{82.4}\\
                $L=20$       &     20.2    &  14.1 & 0.98 & 81.3  & 81.3
            \end{tabular}
        }
    \end{minipage} 
    \hfill
    \begin{minipage}{.49\linewidth}
      \centering
        \caption{Performance of the VQ-MAE-AV model when using the contrastive (1st row), the generative (2nd row), or both (3rd row) loss functions during the self-supervised pre-training.}
        \label{tab:contrastive}
            \begin{tabular}{cccc}
            \hline
                 Contrastive   & Generative & Accuracy \small{(\%)} \\ \hline
                 \checkmark  & \ding{55}  & 75.2  \\
                \ding{55}   & \checkmark  & 84.3  \\
                 \checkmark  & \checkmark  &   \textbf{84.8}
            \end{tabular}
    \end{minipage}
    \hfill \vspace{.5cm}
    \begin{minipage}{0.49\linewidth}
      \centering
        \caption{Performance and number of parameters of the baseline VQ-MAE-AV model (first row, in blue), compared to variations with different modality-fusion strategies at the encoder and decoder (other rows).}
        \label{tab:architecture}
            \begin{tabular}{cccc}
            \hline
                   Encoder  & Decoder & Param. (M)  & Acc. \small{(\%)} \\ \hline
                 \color{blue}\emph{SAF} & \color{blue}\emph{SAF} & \color{blue}13.5 & \color{blue}81.5  \\
                  \emph{CAF} & \emph{SAF} & 25.0 & 82.8  \\
                  \emph{SAF} & \emph{CAF} & 18.4 & 81.8  \\
                 \emph{CAF} & \emph{CAF} & 30.0 & \textbf{83.0}  
            \end{tabular}
    \end{minipage}   
    \hfill 
    \begin{minipage}{0.49\linewidth}
      \centering
        \caption{Performance of the baseline VQ-MAE-AV model (1st row, in blue), compared to variations of the emotion recognition model (other rows).}
        \label{tab:pooling}
            \begin{tabular}{ccc}
            \hline
                Emotion recognition model          & Accuracy \small{(\%)} &  F1 score \small{(\%)}\\ \hline
                \color{blue}Attention Pooling       & \color{blue}81.5 & \color{blue}80.1  \\
                Mean Pooling       & 78.1 & 78.4  \\
                Query2Emo       &  \textbf{84.3} & \textbf{84.8}  
            \end{tabular}
    \end{minipage}  
    \hfill
    \begin{minipage}{.49\linewidth}
      \centering
        \caption{Performance of the baseline VQ-MAE-AV model using the audio and visual modalities (1st row, in blue), compared to the equivalent model using only the audio modality (2nd row) or the visual modality (3rd row).}
        \label{tab:modalities}
            \begin{tabular}{ccccc}
                \hline
                Modality          & Accuracy \small{(\%)} & F1 score \small{(\%)}\\ \hline
                \color{blue}Audio + visual            & \color{blue}\textbf{81.5}  & \color{blue}\textbf{80.1} \\
                Audio           & 73.2 & 72.8\\
                Visual             & 74.1 & 73.9
            \end{tabular}
    \end{minipage} 
\end{table*}

\subsection{Ablation study}
\label{sec:ablation}

In this section, we present a series of ablation experiments using the RAVDESS dataset. Our objective is to evaluate the impact of various hyperparameters and model designs on the emotion recognition performance of the proposed VQ-MAE-AV model. 
In the ablation experiments, we systematically vary one single hyperparameter or model block at a time, starting from a baseline configuration. This baseline uses self-attention fusion for both the encoder and the decoder (see Section~\ref{subsec:encoder_decoder}), it uses only the generative cross-entropy loss function for self-supervised pre-training (see Section~\ref{subsec:loss_function}), and it relies on attention pooling for supervised fine-tuning (see Section~\ref{subsec:finetuning}). This configuration differs from the best-performing one used for the audiovisual emotion recognition experiments presented in Section~\ref{sec:emotion_recognition}, which used the cross-attention fusion strategy for the encoder and decoder, both generative and contrastive loss functions for self-supervised pre-training, and the Query2Emo strategy for supervised fine-tuning. For clarity and ease of interpretation, the baseline configuration is highlighted in blue in the tables of results of the present section.

\subsubsection{Impact of pre-training and fine-tuning}
Table~\ref{tab:finetuning} shows the significance of pre-training and fine-tuning the VQ-MAE-AV model for audiovisual SER. 
Self-supervised pre-training of the model for the unmasking task on the VoxCeleb2 dataset substantially improves the emotion recognition performance, with the accuracy rising from $29.6$\% to $81.5$\%. Fine-tuning the encoder during the supervised training on RAVDESS is also essential, as keeping it frozen leads to a $11$\% drop in accuracy.

\subsubsection{Impact of the encoder depth}
\label{subsubsec:encoder_depth}

Table~\ref{tab:depth} shows the impact of varying the number $L$ of attention blocks in the VQ-MAE-AV encoder, in terms of emotion recognition performance (accuracy and F1 score), number of parameters, the number of floating-point operations (FLOPs) and the runtime (ms) required to make a forward pass in the model for 2 second-long input sequence. The results indicate that increasing the number of blocks in the encoder leads to improved emotion recognition performance up to $L=16$; the accuracy decreases by $1.1$\% for $L=20$. Moreover, it can be seen that this increase in performance comes at the expense of a larger model that requires more FLOPs to compute the prediction.

\subsubsection{Impact of the generative and contrastive loss functions}

Table~\ref{tab:contrastive} shows the impact of using either the generative, the contrastive, or both loss functions during the self-supervised pre-training of the model. This experiment is conducted from the best-performing VQ-MAE-AV model (corresponding to the results in Section~\ref{sec:emotion_recognition}) instead of the baseline one. It can be seen that the best performance is achieved when using the two loss functions. When training VQ-MAE-AV exclusively with the contrastive (resp. generative) loss, the accuracy drops by $9.6$\% (resp. $0.5$\%).

\subsubsection{Impact of the modality-fusion strategies at the encoder and decoder}
Table~\ref{tab:architecture} shows the the performance obtained with the different modality-fusion strategies described in Section~\ref{subsec:encoder_decoder}, including all possible combinations for the VQ-MAE-AV encoder and decoder: \emph{self-attention fusion} for both the encoder and decoder (SAF-SAF); \emph{self-attention fusion} for the encoder and \emph{cross-attention fusion} for the decoder (SAF-CAF); \emph{cross-attention fusion} for the encoder and \emph{self-attention fusion} for the decoder (CAF-SAF); and \emph{cross-attention fusion} for both the encoder and decoder (CAF-CAF). As can be seen, CAF-CAF achieves the highest accuracy with a $1.5$\% improvement over SAF-SAF, followed by CAF-SAF with a $1.3$\% improvement over SAF-SAF, and then SAF-CAF with only a $0.3$\% improvement. The cross-attention fusion architecture at the encoder achieves the best performance, as shown by the CAF-CAF and CAF-SAF configurations. However, there exists a trade-off between performance and the number of model parameters. Notably, CAF-CAF involves slightly more than twice as many parameters as SAF-SAF.

\subsubsection{Impact of the emotion recognition model} Table~\ref{tab:pooling} illustrates the impact of various emotion recognition models used on top of the VQ-MAE-AV encoder for supervised SER. As could be expected, attention pooling and Query2Emo outperform the naive mean pooling strategy. Among the two, Query2Emo performs the best, with an accuracy gain of $2.8$\%. 

\subsubsection{Impact of the modalities} Table~\ref{tab:modalities} compares the performance of the proposed multimodal model with the equivalent model trained using only the audio or visual modalities. It can be seen that exploiting both modalities greatly improves the performance, with accuracy gains of $5.6$\% and $8.3$\% compared to using only the visual and audio modalities, respectively.

\subsubsection{Impact of the audio and visual discrete token size}
\label{sec:impact-token-size}

\begin{figure}[t]
    \centering
    \includegraphics[width=0.48\textwidth]{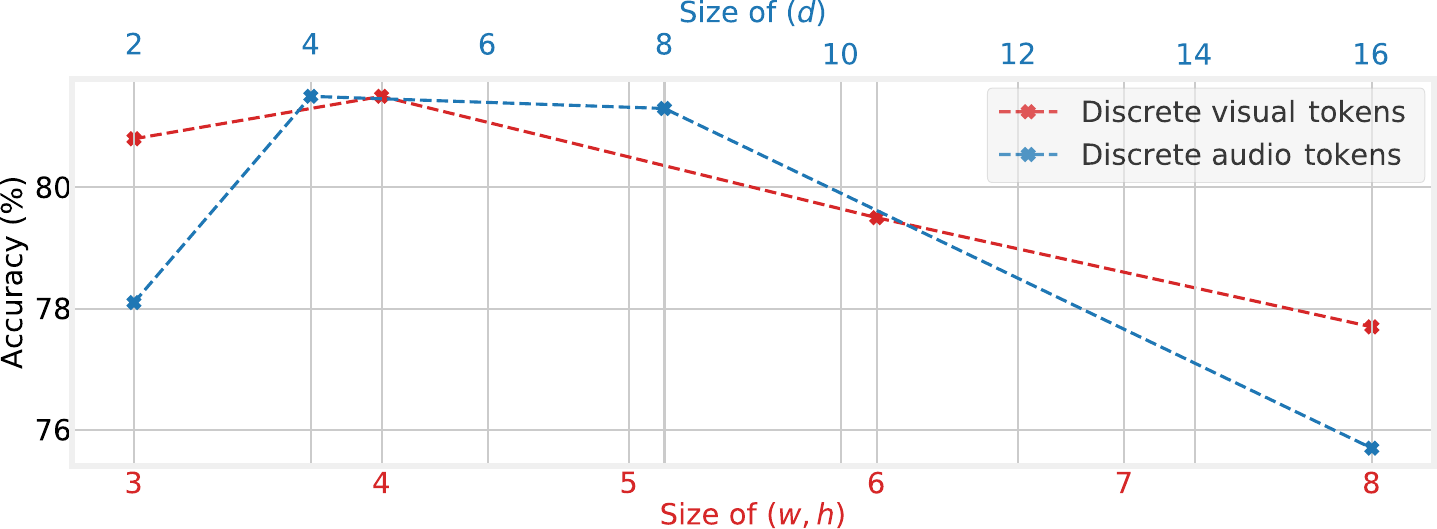}
    \caption{Impact of the discrete audio and visual token size on emotion recognition.}
    \label{fig:size-tokens}
\end{figure}

Fig.~\ref{fig:size-tokens} shows the impact of the dimension of the discrete visual and audio tokens in terms of emotion recognition accuracy. In this figure, $h$ and $w$ represent the horizontal and vertical dimensions of the token in the visual modality, while $d$ represents the frequency dimension of the tokens in the audio modality. Moreover, we only consider the case $h=w$. The results reveal that it is important to select these parameters carefully, as their value significantly impacts the performance. Based on our experiments, we recommend fixing them to ($h=w=4,~d=4$).

\subsubsection{Summary}

The ablation study highlighted key factors that impact SER performance, including attention blocks, model structure, hyperparameters, and training losses. Fine-tuning on emotional datasets proved crucial, as VQ-MAE-AV initially learns general representations through masked token reconstruction, requiring adaptation for SER (see Table~\ref{tab:finetuning}). Integrating both audio and visual modalities further enhanced performance over single-modality inputs (see Table~\ref{tab:modalities}). Additionally, increasing attention blocks in the encoder improves SER performance up to a saturation point (see Table~\ref{tab:depth}). Other factors, such as cross-attention fusion, token size, contrastive learning, and emotion recognition models, offer smaller but notable improvements.

\section{Conclusion}
\label{sec:conclusion}

Masked autoencoding is a versatile self-supervised learning approach that can be adapted to various types of data. This paper introduced the VQ-MAE-AV model for learning representations of audiovisual speech data, which could be extended to other multimodal sequential data.
VQ-MAE-AV took as input a discrete audio representation and a discrete visual representation obtained via two separate VQ-VAEs. These representations were then divided into multiple discrete tokens, with spatio-temporal tokens for the visual modality and spectro-temporal tokens for the audio modality.
Pre-trained on the VoxCeleb2 dataset and fine-tuned on emotional audiovisual speech datasets, the experiments showed that the VQ-MAE-AV model effectively combines the audio and visual modalities for audiovisual SER, outperforming several state-of-the-art methods in both controlled and in-the-wild conditions.

During self-supervised pre-training, the VQ-MAE-AV model learns general-purpose audiovisual speech representations that require fine-tuning for SER. Unlike contrastive SSL approaches, where simple linear probing often yields satisfactory results, our model necessitates fine-tuning on emotional datasets to effectively adapt its representations for emotion recognition. This is a limitation, but it is also an opportunity, as the learned representations could be adapted for various tasks beyond SER \citep{baevski2022data2vec}, such as audiovisual speech recognition, speaker identification, or cross-modal generative tasks, like in AnCoGen \citep{sadok2025ancogen}.

VQ-MAE-AV employs masked modeling as an SSL pre-training paradigm, enabling efficient cross-modal integration. This approach allows the model to learn general audiovisual representations without relying on complex fusion mechanisms, making it adaptable to incorporate additional modalities such as human gesture and pose. This is a particularly interesting avenue for future work on multimodal human behavior analysis, in particular emotion recognition.

\section*{Acknowledgments}
This work was supported by Randstad corporate research chair. It was performed using computational resources from the “M\'esocentre” computing center of Universit\'e Paris-Saclay, CentraleSup\'elec and \'Ecole Normale Sup\'erieure Paris-Saclay supported by CNRS and R\'egion \^Ile-de-France (https://mesocentre.universite-paris-saclay.fr/). 





\bibliographystyle{model2-names}
\bibliography{refs}





\end{document}